# Gbits/s physical-layer stream ciphers based on chaotic light

**Qingchun Zhao and Hongxi Yin[*]**

*Lab of Optical Communications and Photonic Technology, School of Information and Communication Engineering, Dalian University of Technology, Dalian 116023, China*

*\*Corresponding author:* hxyin@dlut.edu.cn

We propose a novel high-speed stream cipher encryption scheme based on the true random key generated by a chaotic semiconductor laser. A 5-Gbits/s non-return-to-zero plaintext is successfully encrypted and decrypted using this cryptography. The scheme can be applied in the areas of real-time high-speed physical encryption.

*OCIS codes:* 060.4785, 140.1540, 200.4560.

With the rapid development of modern information networks and privacy protection, high-speed and secure cryptographic systems are becoming increasingly urgent. Modern cryptography is based on Shannon's theory claiming that "the adversary knows the system", which means that the security of cryptography is mainly relies on the key rather than the encryption algorithm. Hence, the key (i.e. random numbers) plays the essential role in a cipher system.

Up to now, there are two types of keys for modern cryptography, namely, true random numbers (TRNs) and pseudo-random numbers (PRNs). PRNs are generated from a seed utilizing some deterministic algorithms [1]. Although current PRNs have higher generation rate compared with TRNs, the main drawback for PRNs in applications is pseudo-randomness, not true randomness. This shortcoming makes the message be intercepted by eavesdropper. From this point of view, only TRNs are suitable for the one-time pad. Physical random phenomena, such as noise in resistors [2], phase noise of lasers [3], radioactive decay [4], etc., are the favorite entropy sources to generate ideal TRNs. However, the rates for these true random numbers are no more than 20 Mbits/s [5], which extremely limits their applications in modern cryptography. Fortunately, Uchida *et al*. firstly proposed a fast 1.7-Gbits/s true random number generator (TRNG) based on semiconductor lasers operating in chaos [6]. Kanter and his companions improved the TRNG and achieved TRNs with the rates of 12.5 Gbits/s and 300 Gbits/s [7,8].

In this letter, we propose a setup for Gbits/s stream ciphers on the basis of physical devices, which exploits a novel application of TRNs generated by a chaotic semiconductor laser. The numerical simulations confirm that the encryption and decryption processes are feasible. This physical-layer stream ciphers can be transmitted at a high speed to ~ Gbits/s through commercial fiber-optic link.

The detailed schematic diagram for stream ciphers based on the true random key generated by a chaotic semiconductor laser is shown in Fig. 1. The entire setup consists of four blocks: chaotic light source (left red box), key generation (right red box), encryption (green box, Alice), and decryption (green box, Bob). The block of chaotic light source is the same as the generation part of chaos in Ref. [9]. The transmitter laser (TL) is modulated by a 2.35 GHz sinusoidal signal through a bias tee. Chaotic light is generated by external optical feedback from a fiber-optic loop, whose polarization is controlled by a polarization controller (PC). The output power of chaotic light is detected by a photodector (PD). The block of key generation is composed of an analog-to-digital converter (ADC), a clock, and a buffer [7]. The recorded chaotic signal is digitized by the 8-bit ADC triggered by a 1 GHz clock. The buffer stores the digital sequence and shifts one bit to execute difference operation with the unshifted bits. The 5-least significant bits of the difference value are stored as a part of the final TRNs. Accordingly, the rate of TRNs is 5 Gbits/s.

The wavelength division multiplex (WDM) technology is adopted to transmit the key and ciphertext separately. For Alice, she uses the key to execute exclusive-OR (XOR) operation with her message (plaintext). The electronic ciphertext and key are transformed to optical signals by modulating the currents of laser diodes (LDs). LD1 and LD2, working at different wavelengths, are coupled into optical fiber via a wavelength division multiplexer (MUX). After long-haul transmission, the plaintext can be obtained by Bob's decryption devices according to the following steps. First, the optical signals are demultiplexed by a wavelength division demultiplexer (DMUX) and then detected by PDs. Low-pass filters (LPFs) are used to prevent unwanted high-frequency noise. The electronic key and ciphertext can be recovered after sampling and decision. Finally, the plaintext can be obtained by executing XOR operation with Bob's key.

In addition, Alice's solitary LD2 can be modeled by a set of rate equations with the laser electric complex amplitude $E$ and carrier density $N$, respectively, as expressed follows:

$$\frac{dE}{dt} = \frac{1+i\alpha}{2}[\frac{g(N-N_0)}{1+\varepsilon|E|^2} - \frac{1}{\tau_p}]E, \tag{1}$$

$$\frac{dN}{dt} = \frac{I}{qV} - \frac{N}{\tau_N} - \frac{g(N-N_0)}{1+\varepsilon|E|^2}|E|^2, \tag{2}$$

where the central wavelength is 1550 nm. The following parameters were used in the simulations: transparency carrier density $N_0=0.4\times10^6$ μm$^{-3}$, threshold current $I_{th}=12$ mA, differential gain $g=2.125\times10^{-3}$ μm$^3$ns$^{-1}$, carrier lifetime $\tau_N=2$ ns, photon lifetime $\tau_p=2$ ps, round-trip time in laser intra cavity $\tau_{in}=9$ ps, linewidth enhancement factor $\alpha=5.5$, gain saturation parameter $\varepsilon=3\times10^{-5}$ μm$^3$, active layer volume $V=150$ μm$^3$.

In order to generate true random numbers, experimental data of chaotic light are more suitable than numerical data by solving the rate equations of semiconductor laser. The source of chaotic light was provided by Hong and Shore [9], where the detailed experimental conditions are presented. Figure 2 shows a segment of chaotic signal recorded by an oscilloscope with the sampling rate of 10 GHz. The sampling rate of ADC is controlled by a clock whose rate is set to 1 GHz (big red circles in Fig. 2). Hence, only the 1-GHz sampled signal is utilized for key generation.

We adopt a back-to-back configuration without transmission via optical fiber with the aim of simplifying the numerical simulations. Alice's key is shared by Bob to decrypt the ciphertext after sampling and decision.

The generated 5-Gbits/s key, employed for plaintext encryption, is shown in Fig. 3 (a). The plaintext is 5-Gbits/s non-return-to-zero plaintext. As can be seen in Fig. 3 (b), the ciphertext (blue line) successfully masks the plaintext. The output power of LD varies with the ciphertext modulating the current of LD. After PD, LPF, sampling and decision, the plaintext can be decrypted with no error-bits in this back to back cryptography configuration. The eye diagrams for this cryptography are shown in Fig. 4. The eye diagram for signal after sampling is obviously better than that for the output power of LD transmitted in the public channel. Hence, the decryption block maintains Bob's correct plaintext.

In conclusion, we propose a setup for high-speed physical-layer stream ciphers whose key is generated by a chaotic semiconductor laser. The encryption and decryption processes are numerically simulated. The future hot topics of this area are how to use TRNs generated by chaotic LD in block cipher and public key cipher.

This work was supported in part by the National Natural Science Foundation of China (NSFC) under Grant 60772001, Open Fund of State Key Laboratory of Advanced Optical Communication Systems and Networks (Peking University), and Scientific Research Start-up Fund of Dalian University of Technology for Introduced Scholars, China. The authors acknowledge Dr. Yanhua Hong and Prof. K. A. Shore for providing the experimental data of chaotic signal generated by semiconductor lasers.

List of figure captions:

Fig. 1. (Color online) Schematic diagram for high-speed stream cipher based on true random numbers. PC, polarization controller; PD, photodector; LPF, low pass filter; ADC, analog-to-digital converter; TL, transmitter laser; LD, laser diode; MUX, multiplexer; DMUX, demultiplexer .

Fig. 2. (Color online) A 10 ns experimental trace of the TL intensity (small blue dots) recorded at 10GHz, while the sampled points at 1GHz (big red circles).

Fig. 3. (Color online) 100 ns numerical results of encryption and decryption. (a) 5-Gbit/s key; (b) plaintext (blue line, moved up and compressed to guide eyes) and the output power of LD (red line). ; (c) encrypted (blue line) and decrypted (red line, slightly moved up) plaintext.

Fig. 4. (Color online) Eye diagrams for (a) the output power of LD; (b) signal after sampling.

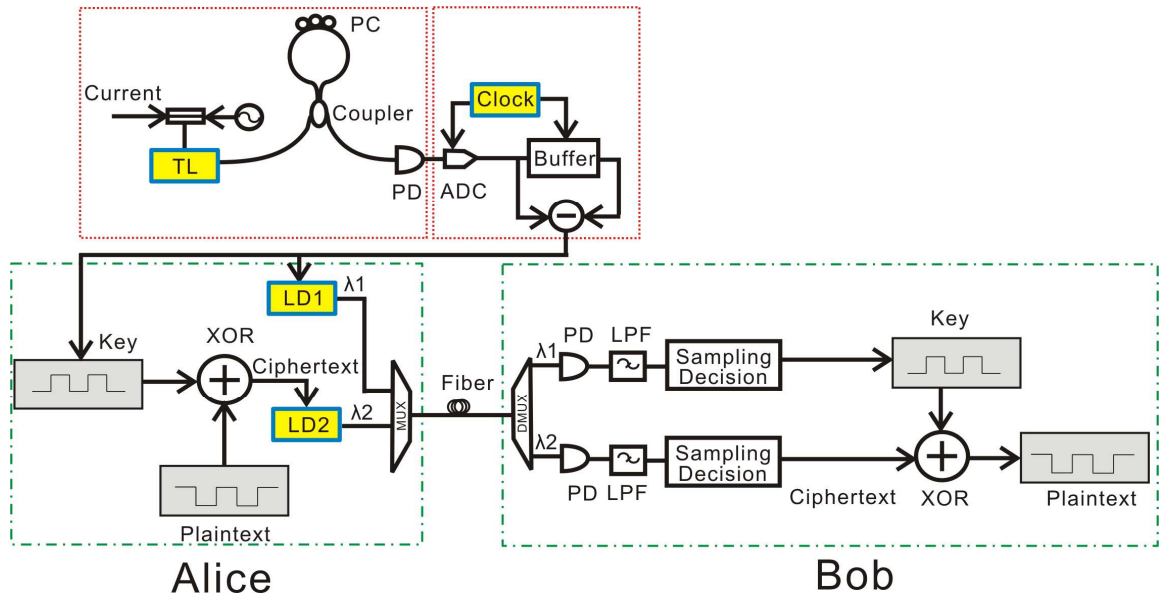

Fig. 1. (Color online) Schematic diagram for high-speed stream cipher based on true random numbers. PC, polarization controller; PD, photodector; LPF, low pass filter; ADC, analog-to-digital converter; TL, transmitter laser; LD, laser diode; MUX, multiplexer; DMUX, demultiplexer .

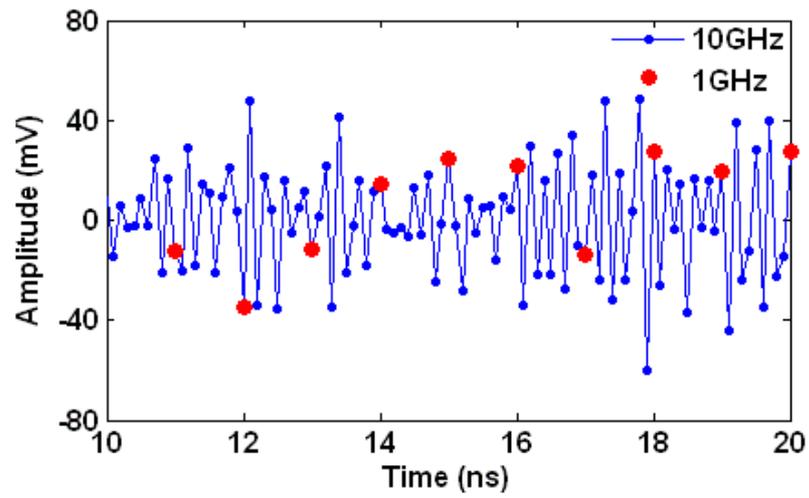

Fig. 2. (Color online) A 10 ns experimental trace of the TL intensity (small blue dots) recorded at 10GHz, while the sampled points at 1GHz (big red circles).

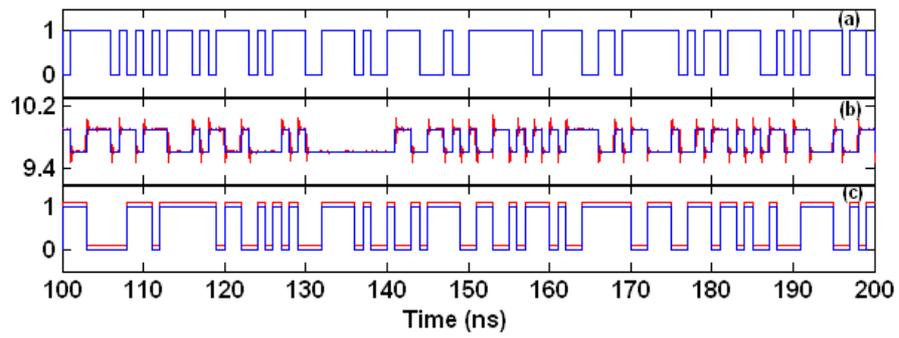

Fig. 3. (Color online) 100 ns numerical results of encryption and decryption. (a) 5-Gbit/s key; (b) plaintext (blue line, moved up and compressed to guide eyes) and the output power of LD (red line). ; (c) encrypted (blue line) and decrypted (red line, slightly moved up) plaintext.

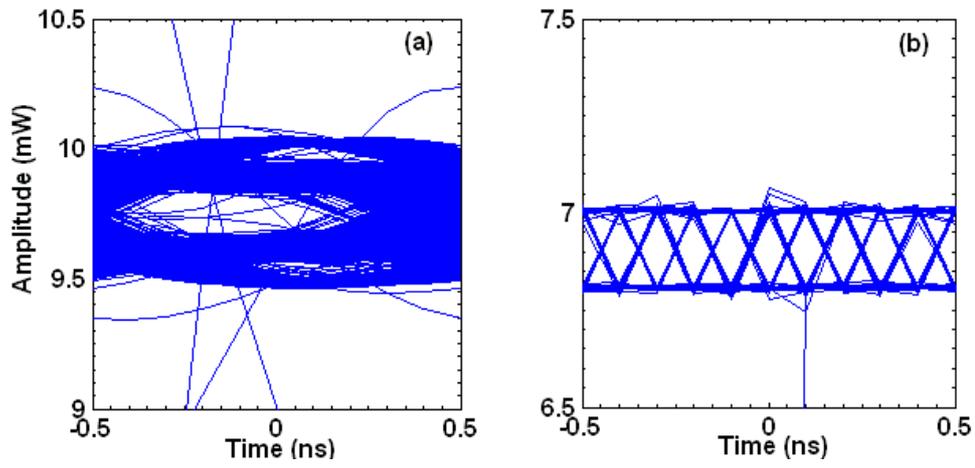

Fig. 4. (Color online) Eye diagrams for (a) the output power of LD; (b) signal after sampling.